# STRUCTURE-SENSITIVE MECHANISM OF NANOGRAPHENE FAILURE

ELENA F. SHEKA[1], NADEZHDA A.POPOVA[1], VERA A.POPOVA[1], EKATERINA A. NIKITINA[1,2], and LANDYSH H. SHAYMARDANOVA[1].

[1]Peoples' Friendship University of Russia, 117198 Moscow

[2]Institute of Applied Mechanics RAS, 119991 Moscow

sheka@icp.ac.ru

**ABSTRACT**: The response of a nanographene sheet to external stresses is considered in terms of a mechanochemical reaction. The quantum chemical realization of the approach is based on a coordinate-of-reaction concept for the purpose of introducing a mechanochemical internal coordinate (MIC) that specifies a deformational mode. The related force of response is calculated as the energy gradient along the MIC, while the atomic configuration is optimized over all of the other coordinates under the MIC constant-pitch elongation. The approach is applied to the benzene molecule and (5, 5) nanographene. A drastic anisotropy in the microscopic behavior of both objects under elongation along a MIC has been observed when the MIC is oriented either along or normally to the C-C bonds chain. Both the anisotropy and high stiffness of the nanographene originate at the response of the benzenoid unit to stress.

Oppositely to real physical experiments, when changing the object shape under loading is usually monitored, computational experiments deal with the total energy response to the object shape deformation that simulates either tension and contraction or bending, screwing, shift, and so forth. As for graphene and carbon nanotubes (CNTs), whose mechanical properties are amenable to experimental study with difficult, the computational experiments takes on great significance.

A lot of works are devoted to the calculation of mechanical properties of nanographenes and CNTs in due course of which two approaches, namely, continuum and atomistic ones have been formulated. The continuum approach is based on the well developed theory of elasticity of continuous solid media applied to shells, plates, beams, rods, and trusses. The latter are structure elements used for the continuum description. When applying to either CNT or nanographene, their lattice molecular structure is presented in terms of the above continuum structure elements and the main task of the calculation is the reformulation of the total energy of the studied atomic-molecular nanocarbon system subjected to changing in shape by that in terms of the continuum structure elements. This procedure involves actually the adaptation of the theory of elasticity of continuous media to nanosize objects which makes allowance for introducing macroscopic basic mechanical parameters such as Young's modulus (E), the Poisson ratio (ν), the potential energy of the elastic deformation, etc into the description of mechanical properties of the nanocarbons under interest. Since the energy of these nanocarbons is usually calculated in the framework of the modern techniques (Monte-Carlo, molecular dynamics, quantum chemistry), which takes the object atom structure into account, the main problem of the continuum approach is a linkage between molecular configuration and continuum structure elements. Nanoscale continuum methods (see Refs. 1-5 and references therein), among which those based on the structural mechanics concept [6] are the most developed, have shown the best ability to simulate nanostructure materials. In view of this concept, CNT and graphene are geometrical frame-like structures where the primary bonds between two nearest-neighboring atoms act like load-bearing beam members, whereas an individual atom acts as the joint of the related beams [7-10].

The basic concept of the atomistic approach consists in obtaining mechanical parameters of the object from results of the direct solutions of either Newton motion laws [10, 11] or Schrödinger equations [12, 13] of a certain object that changes its shape following a particular algorithm of simulation of the wished type of deformation. In this case not energy itself, but forces applied to atoms become the main goal of calculations. These forces are input later into the relations of macroscopic linear theory of elasticity and lay the foundation for the evaluation of micro-macroscopic mechanical parameters such as Yung's modulus ($E^*$), the Poisson ratio ($v^*$), and so on. Nothing to mention that parameters $E$ and $E^*$ as well as $v$ and $v^*$ are not the same so that their coincidence is quite accidental. Obviously, atomistic approach falls in opinion comparing with the continuum one due to time consuming calculations and, as a result, applicability to smaller objects. However, it possesses doubtless advantages concerning the description of the mechanical behavior of the object under certain loading (shape changing) as well as exhibiting the deformation and failure process at atomic level.

Following the wish to emphasize the latter aspect of the atomistic approach, we suggest in the current paper to go beyond a conventional energy-strain-response concept and to consider the mechanism of the tensile deformation leading to the failure and rupture of a nanographene sheet as the occurrence of a mechanochemical reaction. A similarity between mechanically induced reaction and the first-type chemical ones, first pointed out by Tobolski and Eyring more than sixty years ago [14], suggested the use of a well developed quantum-chemical (QCh) approach of the reaction coordinate [15] in the study of atomic structure transformation under deformation. Firstly applied to the deformation of poly(dimethylsiloxan) oligomers [16], the approach has revealed a high efficacy in disclosing the mechanism of failure and rupture of the considered polymers.

The main point of the approach concerns the reaction coordinate defining. When dealing with chemical reactions, that is usually selected among the internal ones (valence bond, bond angle or torsion angle) or is presented as a linear combination of them. Similarly, *mechanochemical internal coordinates* (MICs) were introduced as modified internal coordinates defined in such a way as to be able to specify the considered deformational modes [16, 17]. The MICs thus designed are to meet the following requirements:

1. Every MIC is a classifying mark of a deformational mode: uniaxial tension and contraction are described by linear MICs similar to valence bonds, bending is characterised by a MIC similar to valence angle, and screwing is attributed to MICs similar to torsional angles. Thus introduced MICs are microscopic analogues of macroscopic beam elements of structural mechanics [6].
2. Every MIC is determined in much the same way as the other internal coordinates except for a set of specifically selected support atoms.
3. The MIC relevant to a particular deformational mode is excluded from the QCh optimisation procedure when seeking the minimum of the total energy.
4. A *force of response* is determined as the residual gradient of the total energy along the selected MIC. This logic is dictated by the general architecture of the conventional QCh software where the force calculation, namely, the total energy gradient calculation, is the key procedure.

Implementation of the MIC concept in the framework of DYQUAMECH software [18], which is based on the Hartree-Fock unrestricted version of the CLUSTER-Z1 codes exploiting advanced semiempirical QCh methods [19], provides (i) the MIC input algorithm, (ii) the computation of the total energy gradients both in the Cartesian and internal coordinates, and (iii) the optimization performance in the internal coordinates. Additionally, the program retains all features of the broken symmetry approach, particularly important for odd electronic systems of CNTs [20] and graphene [21].

*Force of response calculation.* The forces, which are the first derivatives of the electron energy $E(R)$ over the Cartesian atom coordinates $R$, can be determined as [16]

$$\frac{dE}{dR} = <\phi\left|\frac{\partial H}{\partial R}\right|\phi> + 2<\frac{\partial \phi}{\partial R}|H|\phi> + 2<\frac{\partial \phi}{\partial P}|H|\phi>\frac{dP}{dR} \quad . \tag{1}$$

Here $\phi$ is the electron wave function of the ground state at fixed nucleus positions, $H$ represents the adiabatic electron Hamiltonian, and $P$ is the nucleus pulse. When calculating (1), a quite efficient computational technique suggested by Pulay [22] was applied. When the force calculation is completed, the gradients are re-determined in the system of internal coordinates in order to proceed further in seeking the total energy minimum by atomic structure optimization. The DYQUAMECH algorithm of the force determination concerns forces applied to each of $i$ MICs. These partial forces $F_i$ are used afterwards for determining all wished micro-macroscopic mechanical characteristics among which we choose the following related to uniaxial tension:

- Force of response $F$: $$F = \sum_i F_i \qquad (2)$$
- Stress $\sigma$: $$\sigma = F/S = \left(\sum_i F_i\right)/S, \qquad (3)$$
  where $S$ is the loading area,
- Young's modulus $E^*$: $$E^* = \sigma/\varepsilon, \qquad (4)$$
  where $\varepsilon = \Delta L_i / L_0$ is the strain and $\Delta L_i$ is the elongation of the $i$-th MIC and is identical to all MICs in the current experiment,
- Stiffness coefficient $k$: $$k = F/\Delta L_i. \qquad (5)$$

The following characteristics are thus obtained when a computational cycle is completed.

1. The atomic structure of the loaded body at any stage of the deformation including bond scission and post-breaking relaxation. Post-breaking fragments can be easily analysed therewith by specifying them as the products of either homolytic or heterolytic reaction. Additional characteristic of the fragments is given in terms of the total $N_D$ and atomically partitioned $N_{DA}$ numbers of effectively unpaired electrons [23] that show a measure of the fragment radicalization and disclose the chemical reactivity of the fragment atoms.

2. A complete set of dynamic characteristics of the deformation that are expressed in terms of microscopic and micro-macroscopic characteristics. The former concern energy-elongation, force-elongation, $N_D$ and/or $N_{DA}$– elongation dependencies that exhibit mechanical behaviour of the object at all stages of the deformation considered at the atomic level. The latter involve stress-strain interrelations in terms of Eqs. (3)-(5) which allow for introducing convenient mechanic characteristics similar to those of the elasticity theory.

In what follows a selected set of the above-listed topics will be presented with respect to tensile deformation of (5, 5) nanographene. The calculations were performed by using AM1 version of the DYQUAMECH program.

*Graphene as the object of mechanical deformation.* From the mechanical viewpoint, benzenoid hexagon structure of graphene put two questions from the very beginning concerning (1) mechanical properties of the benzenoid unit itself and its mechanical isotropy, in particular, and (2) the influence of the units packing on the mechanical properties of graphene as a whole. Conventionally, the mechanic properties of benzenoid units are considered as completely isotropic due to high symmetry of the unit structure [1-4]. However, as shown recently [21] the unit exact symmetry in real nanographenes is much lowed than $D_{6h}$ so that the suggestion of its mechanical isotropy is rather questionable. Moreover, a conclusion about mechanical isotropy does not follow from structural symmetry of the object since the object rupture is connected with the scission of particular chemical bonds whose choice is determined by the configuration of the relevant MICs that suit the geometry of applied loading. That is why the surrounding of thus chosen chemical bonds may be different under deformation in different directions even in high symmetry structures. To check the prediction let us look at the tensile deformation of the benzene molecule subjected to two modes of the uniaxial tension nominated as *zig-zag* (*zg*, along C-C bond) and *arm chair* (*ach*, normal to C-C bond) ones.

*Tensile deformation and rupture of the benzene molecule.* Configurations of two MICs related to each of the two deformation modes are shown in Fig.1. The deformation proceeds as a stepwise elongation of the MICs with the increment $\delta L$=0.05Å at each step so that the current MIC length constitutes $L = L_0 + n\delta L$, where $L_0$ is the initial length of the MIC and $n$ counts the number of the deformation steps. One end of each MIC is fixed (on atoms 1,2 and 1,5 in the case of *ach* and *zg* modes, respectively). Consequently, these atoms are immobilized while atoms 5,6 and 2,6 move along the arrows providing the MIC successive elongation but do not participate therewith in the optimization procedure at each elongation step.

Figure 2 presents the elongation response dependencies, related to the *ach* and *zg* deformation modes, of the total response force $F$ in terms of Eq. (2) (Fig.2*a*), the total $N_D$ (Fig. 2*b*) and partial $N_{DA}$ (Figs. 2*c* and 2*d*) numbers of unpaired electrons that are the main microscopic characteristics of the mechanochemical reaction describing the benzene molecule failure obtained directly from the QCh calculations. As seen in Fig.2, the mechanical behavior of the molecule is highly anisotropic. The force-elongation dependence shown in Fig. 2*a* differs both in the initial linear region and on the final steps exhibiting a considerable enlarging of the failure zone in the case of the *zg* mode in comparison with the *ach* one. Linear elastic behavior is highly restricted and is limited to one-two first deformation steps.

Even more radical difference is illustrated in Figs. 2*b*-2*d* pointing to the difference in the electronic processes that accompany the molecule failure. Obviously, these features are connected with the difference in the MIC atomic compositions related to the two modes, which results in the difference of the structure of the molecular fragments formed under rupture. In the case of the *zg* mode, two MICs are aligned along $C_1$-$C_6$ and $C_3$-$C_4$ molecular bonds and two atomically identical three-atom fragments are formed under rupture. In due course of the *zg* mode, the MIC elongation is immediately transformed into the bond elongation. The C-C bond length of the unstrained benzene molecule of 1.395Å is just a bordering value exceeding which violates a complete covalent coupling of the molecule odd electrons of two neighbor carbon atoms and causes the appearance of effectively unpaired electrons [20, 21]. That is why the increment value of 0.05Å is significant enough for the unpaired electrons appearance even at the first step of elongation. The bonds breaking occurs when the elongation achieves 0.2-0.3Å (these very values determine the maximum position of the force-elongation dependence in Fig.2*a*) but the two three-atom radicals are stabilized only when the elongation exceeds 1.2Å (see Fig.2*d*).

In the case of the *ach* mode, the corresponding MICs connect atoms 1&5 and 2&6, respectively so that only ~40% of the MIC elongation is transformed into that of each of two C-C bonds that rest on the MIC. This explains why $N_{DA}$ values on all carbon atoms are quite small in this case (Fig.2*c*) until the MIC elongation $\Delta L$ is enough to provide the bond breaking. Actually, the bond breaking is not a one-moment process and, as seen in Fig.2*a*, it is originated at $\Delta L$=0.3 Å and is completed at $\Delta L$=0.6 Å. Two-atom and four-atom fragments are formed when the molecule is broken. At the rupture moment, the former is a stretched acetylene molecule, which is followed by the presence of unpaired electrons on atoms 5 and 6 (see Fig.2*c*) due to exceeding the C-C bond length a critical value over which the covalent coupling of the odd electrons is incomplete. However, a further relaxation of the molecule structure at larger elongation shortens the bond putting it below the critical value and unpaired electrons disappear. The second fragment is a biradical whose structure is stabilized at $\Delta L$=0.9 Å.

Therefore, the *zg* and *ach* modes of the tensile deformation of the benzene molecule occur as absolutely different mechanochemical reactions. Besides the difference in the microscopic behavior, the two modes are characterized by different mechanical parameters in terms of Eqs. (2)-(5), whose values are given in Table 1. Taking together, the obtained results evidence a sharp mechanical anisotropy of the benzene molecule in regards the direction of the load application as well as an extremely high stiffness of the molecule. Actually, the considered deformational modes of the benzene molecule do not reproduce exactly the similar situations for benzenoid units in CNTs and graphene, however, the *zig zag* and *arm chair* edge-structure dependence of the mechanical behavior of both CNTs [2, 24] and graphene [4, 25] is evidently provided by the mechanical anisotropy of the benzene molecule to a great extent.

*Tensile deformation of (5, 5) nanographene*. The benzenoid pattern of graphene sheets and a regular packing of the units predetermine the choice of either parallel or normal MICs orientation to the chain of C-C bonds similarly to those introduced for studying the benzene molecule. In the rectangular nanographene sheets and nanoribbons the former orientation corresponds to tensile deformation applied to the zigzag edges while the latter should be attributed to the arm chair edges. We shall nominate them *zg* and *ach* as was done previously. The MIC configurations of the *ach* and *zg* tensile modes of the (5,5) nanographene sheet are presented in Fig. 3. The computational procedure was fully identical to described above for the benzene molecule with the only difference concerning the step increment $\delta L=0.1$Å. The loading area is determined as $S = DL_{0\_z(a)}$, where $D$ is the van der Waals diameter of the carbon atom of 3.35Å and $L_{0\_z(a)}$ is the initial length of the MICs in the case of *zg* and *ach* modes, respectively.

*ach Mode of the graphene tensile deformation*. Figure 4 presents structure images of the selected set of deformation steps. The sheet is uniformally stretched in due course of first 14 steps and breaking the first C-C bond occurs at the 15th step. The breaking is completed at the 17$^{th}$ step and the final structure transformation looks like that occurring at similar deformational mode of benzene: the sheet is divided in two fragments, one of which is a shortened (4, 5) equilibrated nanographene while the other presents a polymerized chain of acetylene molecules transferred into the carbine C=C bond chain.

*zg Mode of the graphene tensile deformation*. Exhibited mechanical anisotropy of the benzene molecule has given grounds to expect the difference in the mechanical behavior of the *ach* and *zg* tensile modes of the studied nanographene but the obtained results has greatly surpassed the all expectations. Figure 5 presents the structure image of the selected set of successive deformation steps revealing a grandiose picture of a peculiar failure of the sheet with so drastic difference in details when comparing to the *ach* mode that only a simplified analogy can assist in a concise description of the picture. The failure of tricotage seems to be a proper model. Actually, as known, the toughness of a tricotage sheet as well as the manner of its failure depends on the direction of the applied stress and the space configuration of its stitch packing. On this language, each benzenoid unit presents a stitch and in the case of the *ach* mode, the sheet rupture has both commenced and completed by the rupture of a single stitch row. In the case of the *zg* mode, the rupture of one stitch is 'tugging at thread' the other stitches that are replaced by still elongated one-atom chain of carbon atoms. Obviously, this difference in the mechanical behavior is connected with the benzenoid packing. Thus, if at the *ach* deformation mode the stitch rupture is connected with the scission of two C-C bonds localized within one benzenoid unit, the breaking of each C-C bond at the *zg* deformation mode touches three benzenoid units promoting the dissolution of the stitch chain in this case.

*A comparative study of the ach and zg modes of the graphene tensile deformation*. The difference in the structural pattern of the two deformation modes naturally leads to the difference in quantitative characteristic of the mechanical behavior in these two cases. Figure 6 shows the total forces of response and the total number of effectively unpaired electrons for both deformation modes versus the elongation. Oppositely to the benzene molecule with no unpaired electrons in the unstrained ground state, the $N_D$ value of unstrained nanographene is quite big due to the elongation of the C-C bond of the unstrained benzenoid unit in comparison to those of the benzene molecule [21]. Figures 6*b* and 6*d* disclose the additional effect of the C-C bond elongation caused by the tensile deformation on the number of effectively unpaired electrons. Since the calculations allow for partitioning the $N_D$ value over atoms, a scrupulous analysis of each bond behavior becomes possible. A detailed consideration of the topic will be presented in details later on.

As seen in the figure, the microscopic behavior of the nanographene at the first stage of the deformation is similar to that of the benzene molecule in both cases since it is connected with the scission of the first C-C bond. As seen from Table 1, the mechanical characteristics determined on the basis of the data related to the first stage of deformation behave similarly to those of the benzene molecule indicating a deep connection of the sheet behavior with that of individual benzenoid unit. At the same time, all micro-macroscopic values are quite different from those of the benzene just pointing to the influence of the unit packing in the sheet.

While the *ach* deformation is one-stage and is terminated at the $20^{th}$ step, the *zg* deformation is multi-stage and proceeds up to $250^{th}$ step followed by the saw-tooth shape of the force-elongation response reflecting a successive stitch dissolution, which is clearly seen in Fig.5 that presents structures related to the steps corresponding to the teeth maxima. And only the one-atom chain cracking at the $249^{th}$ step completes the nanographene rupture. The revealed tendency in the formation of the one-atom chains under stress has been recently experimentally recorded [26, 27]. It seems reasonable to suggest that just the stress originated in due course of the electron beam bombardment of the graphene body causes so peculiar structure transformation calling to life one-atom chains of carbons.

The $N_D$ dependencies shown in Figs. 6*b* and 6*d* clearly evidence the difference in the mechanochemical reactions related to the *ach* and *zg* deformation modes even on the first stage. When the reaction is terminated after achieving the $21^{st}$ step in the former case, the *zg* reaction proceeds further with the $N_D$ dependence looking like a saw-tooth one characteristic for the response force in Fig.6*c*. Important to note, that the discussed $N_D$ dependencies reflect a quantitative changing in the chemical reactivity of the nanographene sheet under tensile deformation.

*Tensile deformation of the hydrogen-terminated (5, 5) nanographene*. The termination of the nanographene sheet edges by hydrogen atoms, preserving a qualitative picture of the nanographene mechanical behavior in general, drastically changes in details. Figure 7 presents the final structures related to the *ach* and *zg* deformation modes that are accomplished on the $20^{th}$ and $122^{th}$ steps, respectively, alongside with the calculated forces of response. In the case of the *ach* mode, the deformation occurs in one stage similarly to that for non-terminated graphene with the only difference concerning the structure of the final fragments that is influenced by the presence of hydrogen atoms. Oppositely to this, the *zg* mode, once multi-stage as previously, proceeds nevertheless quite differently in regards the non-terminated graphene. The first broken C-C bond is located just in the center of the sheet so that further bond scission has led to the formation of a closed one-atom chain of carbon atoms whose splitting as a whole (see Fig.7*b*) from the nanographene edge manifests the termination of the sheet failure. Although this action occurred at the $122^{th}$ step, at far longer elongation than that of the *ach* mode, the latter is twice less than that of the non-terminated graphene.

Table 1 collects the main parameters related to the first stage of deformation. As seen from the table, within this stage, there is an obviously similarity in the behavior of both non-terminated and H-terminated sheets under both deformation modes. The Young moduli data are within the range determined by other calculations [11] and correlate with 1 TPa offered on the basis of experimental observations [28]. However, it should be noted that the region of elongation which supports the linear elastic law, is rather short in all cases, so that the graphene deformation from the very beginning is nonlinear and non-elastic. This circumstance puts the first question about the possibility of the graphene deformation description on the basis of the theory of elasticity by using such parameters as Young moduli, Poisson coefficients, and so forth. The second question follows from the fact that one-stage deformation occurs in the case of the *ach* mode only. Oppositely, *zg* deformation is multi-stage, which puts it outside the theory of elasticity at all and deprives the parameters of any sense. At the same time, the data related to the first stage, once the largest, scale the loading to be applied to graphene for its failure to be occurred.

The mechanochemical-reaction QCh approach, applied to the consideration of the graphene deformation for the first time, has shown a high efficacy in disclosing atomically matched peculiarities that accompany the deformation-failure-rupture process just highlighting the mechanism of the mechanical behavior of the object under particular mode of deformation. The approach has revealed as well that the modern versions of the theory of elasticity adapted to the description of the graphene deformation should be seriously reexamined.

Table 1. Micro-macroscopic mechanical characteristics of benzene molecule and (5, 5) nanographene[1].

| | Critical elongation *nm* | Critical response force $F_{cr}$ N*$10^{-9}$ | Stress σ $(N/m^2)$*$10^9$ | Stiffness coefficient *k* N/m | Young's modulus E* TPa |
|---|---|---|---|---|---|
| Benzene *zg* mode | 0.7 | 12.54 | 153.72 | 17.92 | 3.2 |
| Benzene *ach* mode | 0.7 | 9.82 | 206.31 | 14.03 | 4.23 |
| (5, 5) graphene *zg* mode | 1.6 | 45.08 | 116.18 | 28.17 | 1.14 |
| (5, 5) graphene *ach* mode | 1.6 | 50.09 | 129.11 | 31.31 | 1.2 |
| H-terminated (5, 5) graphene *zg* mode | 1.5 | 47.29 | 121.89 | 31.53 | 1.33 |
| H-terminated (5, 5) graphene *ach* mode | 1.9 | 54.53 | 140.53 | 28.7 | 1.24 |

1) Stress and stiffness coefficient are determined at critical force of deformation. The Young moduli are determined as tangents of slope angles of the stress-strain curves at first steps of deformation.

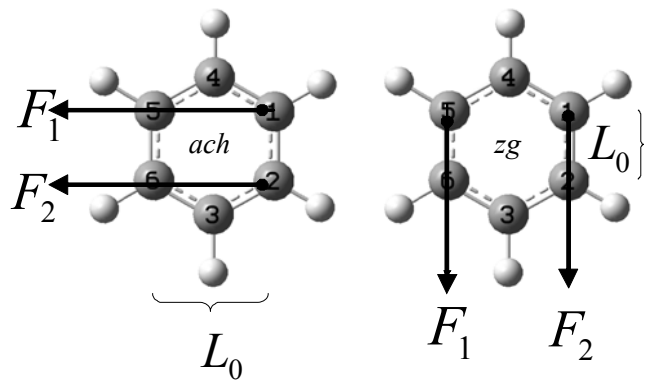

**Figure 1.** Two MICs of uniaxial tension of the benzene molecule for the *ach* and *zg* deformational modes. $L_0$ shows the initial length of the MICs while $F_1$ and $F_2$ number the corresponding forces of response.

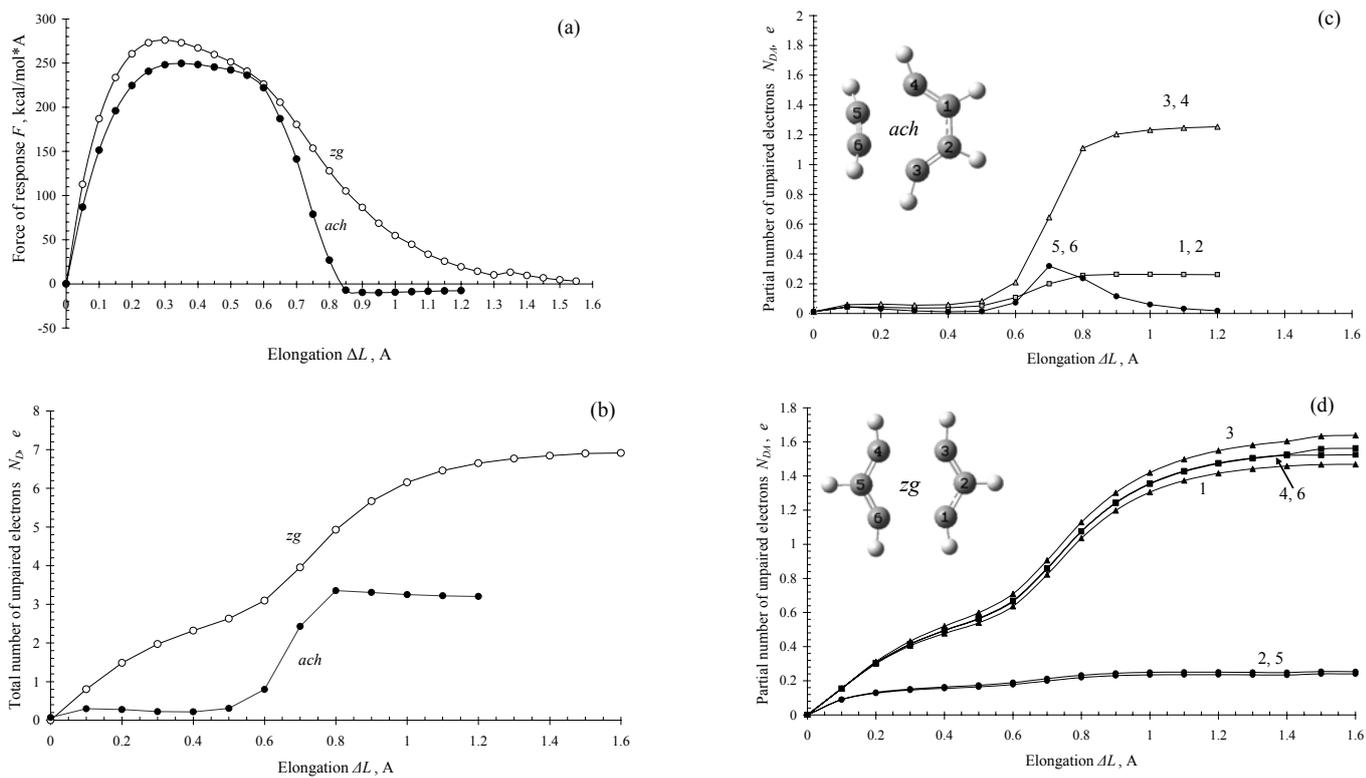

**Figure 2.** Microscopic characteristics of the benzene molecule deformation. Figures in panels *c* and *d* correspond to the atom number on the inserted structures and mark curves related to the relevant atoms.

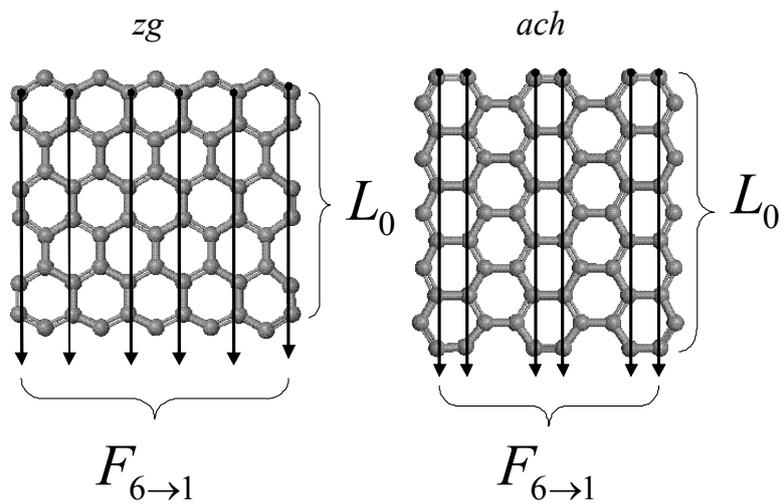

**Figure 3.** Six MICs of uniaxial tension of the (5, 5) nanographene for the *ach* and *zg* deformational modes. $L_0$ shows the initial length of the MICs while $F_{6\to1}$ points six corresponding forces of response.

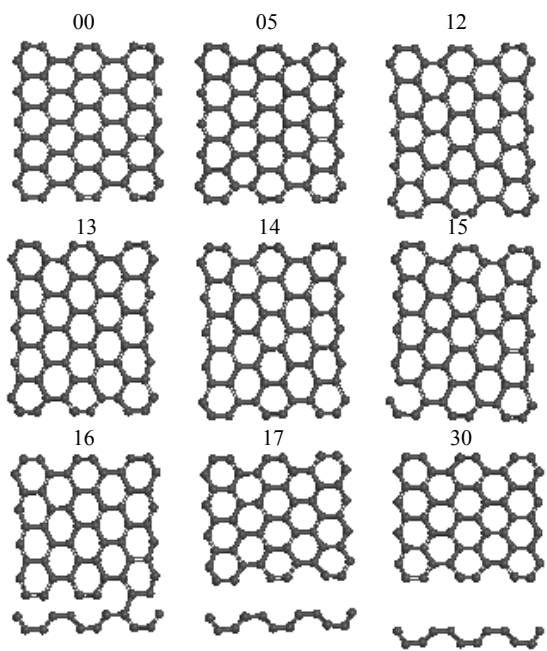

**Figure 4.** Structures of the (5, 5) nanographene under successive steps of the *ach* regime of the deformation. Figures point the step numbers.

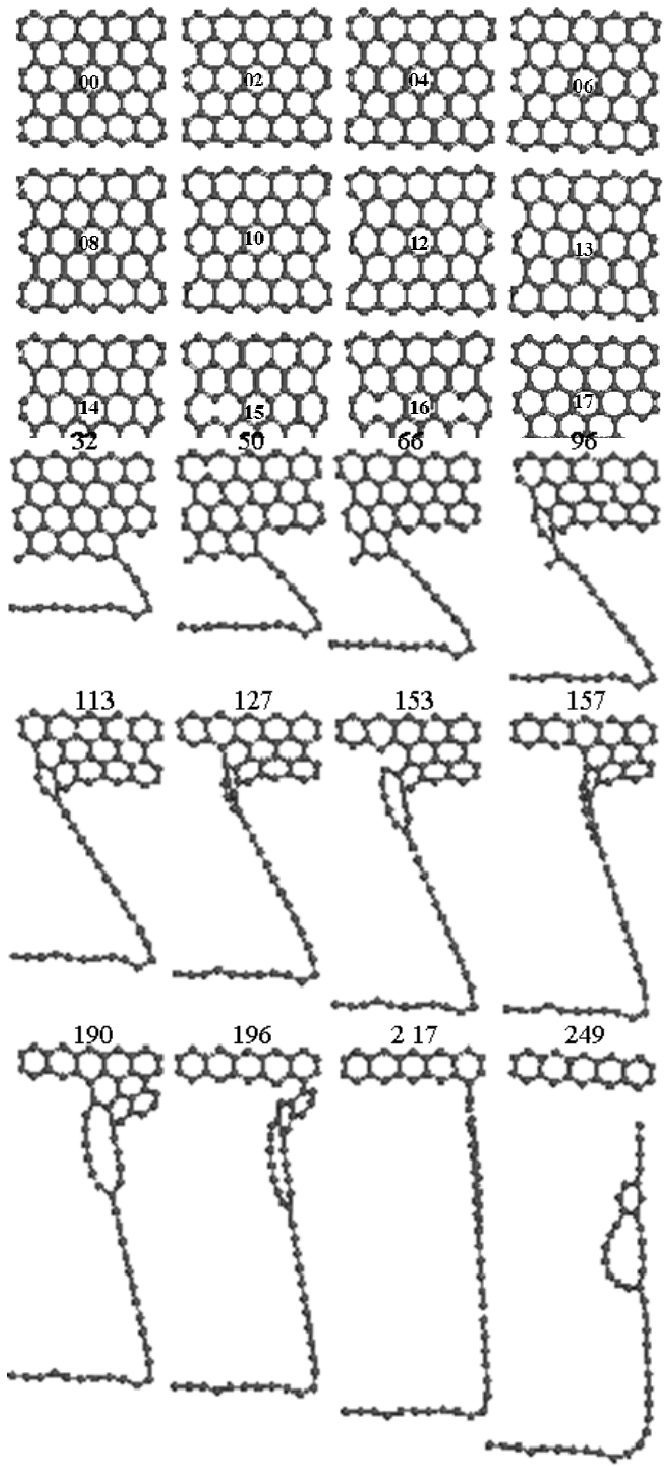

**Figure 5.** Structures of the (5, 5) nanographene under successive steps of the *zg* regime of the deformation. Figures point the step numbers.

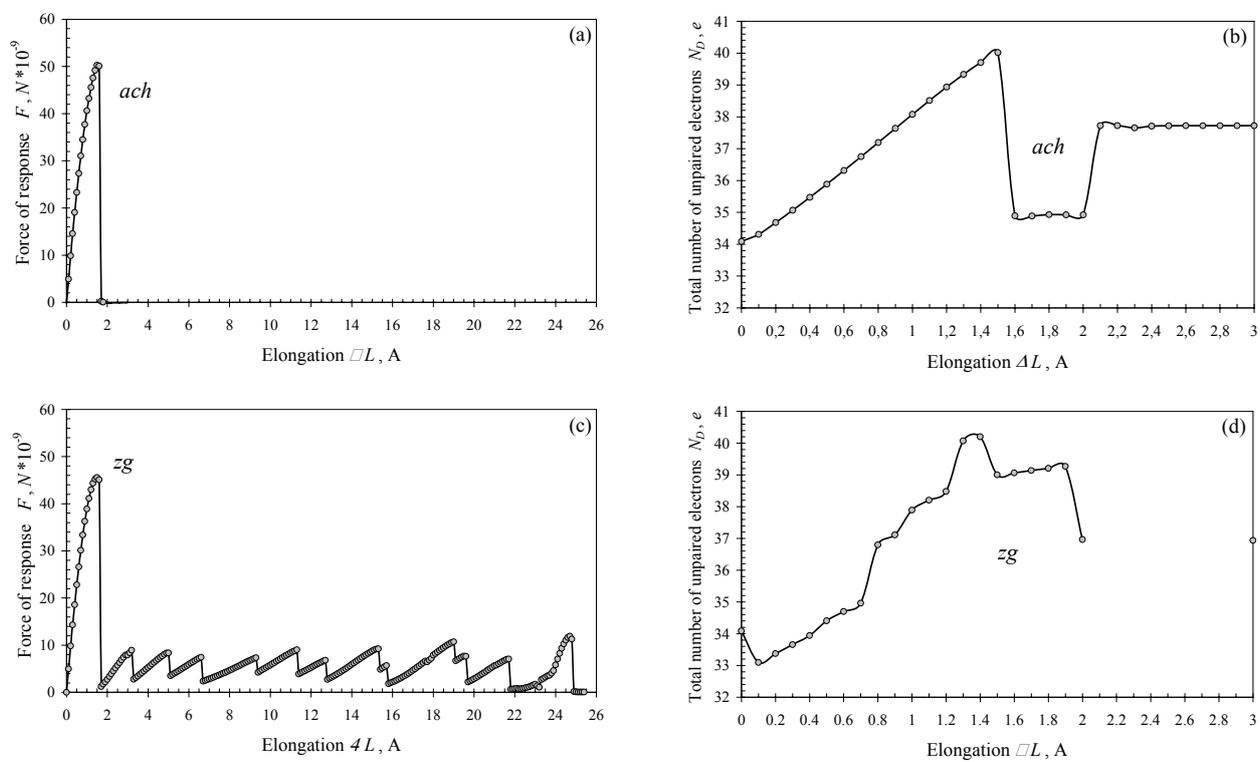

**Figure 6.** Microscopic characteristics of the (5, 5) nanographene deformation.

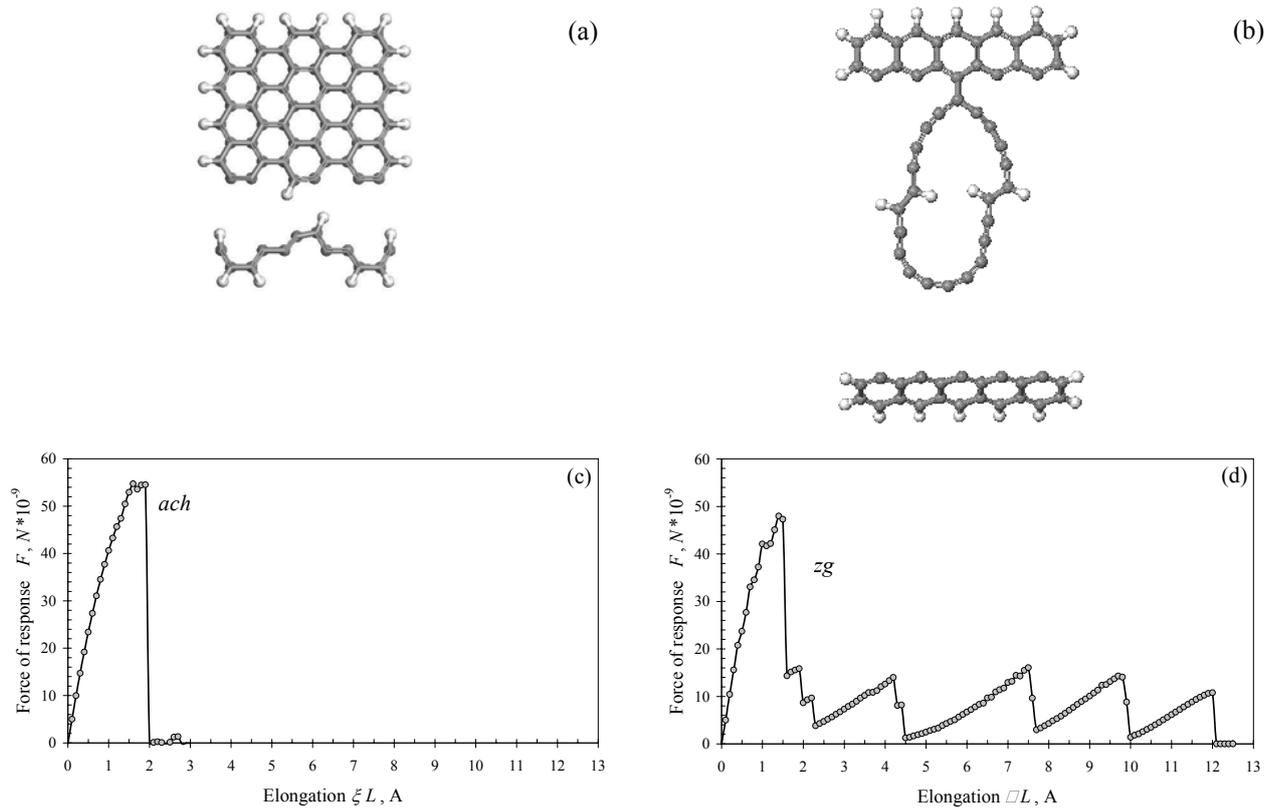

**Figure 7.** Structures of the H-terminated (5, 5) nanographene at the breaking step of the *ach* (*n*=20) (*a*) and *zg* (*n*=122) (*b*) deformation. Forces of response versus elongation for the *ach* (*c*) and *zg* (*d*) modes.